\documentclass[a4paper,11pt]{article}
\usepackage{pos}
\usepackage{graphicx}

\title{Baryon-to-Meson Ratios in Jets from Au+Au and $p$+$p$ Collisions at $\sqrt{s_{\mathrm{NN}}} = 200$ GeV}
\ShortTitle{Baryon-to-Meson Ratios in Jets}

\author*[a]{Gabriel Dale-Gau, on behalf of the STAR collaboration}

\affiliation[a]{University of Illinois at Chicago,\\
  845 W Taylor St, Chicago, IL, USA}

\emailAdd{gdaleg2@uic.edu}

\abstract{Measurements at RHIC and the LHC show strongly enhanced baryon-to-meson yield ratios at intermediate transverse momenta ($p_{\rm{T}}$) in high-energy nuclear collisions compared to $p$+$p$ baseline. This enhancement is attributed to the following QGP effects: strong hydrodynamic flow and parton recombination. Jet probes have been used extensively to gain insights into QGP properties, with substantial modifications to jet yields and internal structures seen across multiple measurements. Despite apparent medium-induced changes to jet fragmentation patterns, the LHC results indicate that in-jet baryon-to-meson ratios remain similar to that of $p$+$p$ measurements and are significantly different from that of the QGP bulk. To explore this behavior at RHIC, we employ particle identification through time-of-flight and TPC $dE/dx$ information alongside jet-track correlations to measure in-jet particle ratios for $p_{\rm{T}}  < 5.0$ GeV/$c$. We present the first in-cone baryon-to-meson yield ratios associated with fully reconstructed jets from Au+Au and $p$+$p$ collisions at $\sqrt{s_{\mathrm{NN}}} = 200$ GeV using the STAR detector at RHIC. }

\FullConference{HardProbes2023\\
 26-31 March 2023\\
 Aschaffenburg, Germany\\}

\begin{document}
\maketitle

\section{Motivation}

Heavy-ion collisions provide a unique environment to study an exotic phase of matter, Quark-Gluon Plasma (QGP), in the laboratory setting. The existence of such a phase has been demonstrated in many ways, for example, through observation of collective partonic flow~\cite{flow}. QGP properties can often be studied by comparing heavy-ion collisions to $p$+$p$ collisions, in which QGP is not expected to be formed. Key signatures of QGP observed through such comparisons include significant modification of charged particle spectra, enhancement of relative baryon to meson production, and jet quenching. The observed differences in these observables demonstrate that the medium impacts charged particle production~\cite{RAA}. Jet quenching-related phenomena can also be studied through hadron production measurements at high transverse momenta or with fully reconstructed jets. Jets, collimated collections of particles produced by fragmentation and hadronization of hard-scattered partons, are present in both $p$+$p$ and heavy-ion collisions. This makes them ideal in-situ probes to study QGP, as we can observe the differences in jet properties with and without the presence of the medium~\cite{JetQuench}. \\
\indent In contrast with elementary collision systems, hard scattering is not the dominant source of particle production in heavy-ion collisions at intermediate transverse momenta, as demonstrated by an enhancement in baryon production relative to meson production~\cite{spectra}.
This enhancement is attributed to the coalescence of partons from the medium. It remains unknown quantitatively to what extent the hard-scattered parton traversing the medium contributes to in-medium coalescence and if the QGP presence modifies the particle composition of the jet shower.

\indent This analysis simultaneously explores two known features of QGP: baryon enhancement and jet modification. The analysis aims to further understanding of how jets fragment in medium as well as QGP hadronization mechanisms. Recent AMPT simulations predict an enhanced in-jet baryon to meson ratio for heavy-ion collision simulations~\cite{ampt}. Using the STAR detector at RHIC, we combine the jet-track correlation technique and particle identification afforded by STAR subsystems to measure the in-jet baryon-to-meson ratios in Au+Au and $p$+$p$ collisions at $\sqrt{s_{\mathrm{NN}}} = 200$ GeV.

\section{Methods}

\begin{figure}
  \includegraphics[width=\linewidth]{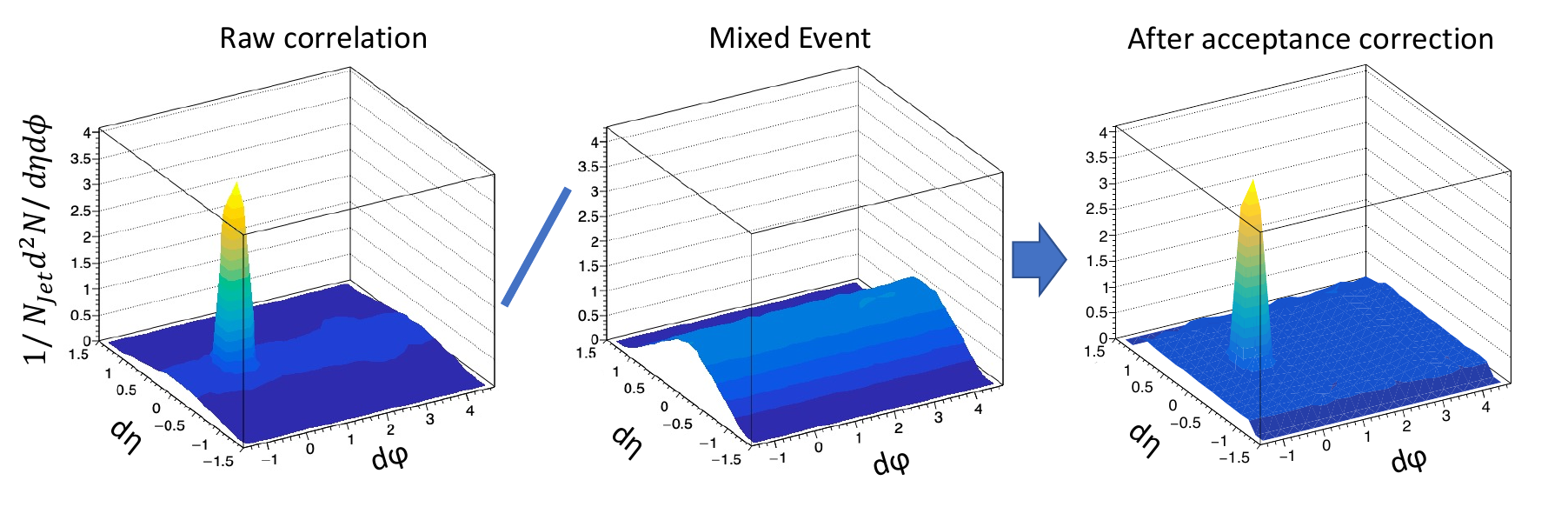}
  \caption{Jet-track correlations in $d\phi$ and $d\eta$. On the left, a raw correlation is shown. In the center, a correlation resulting from a mixed-event procedure. On the right, a pair-acceptance corrected correlation distribution, resulting from dividing the signal by the mixed event correction.}
  \label{accept}
\end{figure}

\begin{figure}
  \includegraphics[width=\linewidth]{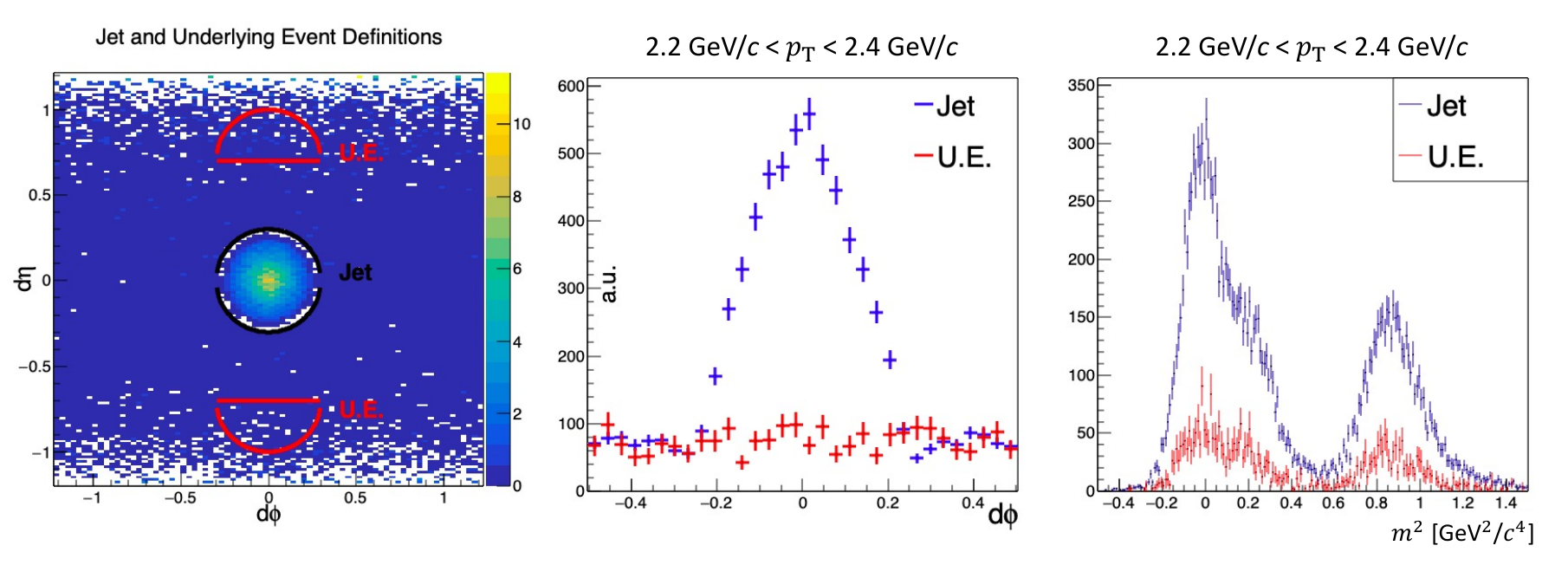}
  \caption{Underlying event subtraction. The left plot shows the areas in $d\phi$ and $d\eta$ selected as jet and underlying event. The center plot shows the $d\phi$ distributions of the two regions. The right plot shows the same two regions overlayed in $m^{2}$ to demonstrate the application of this technique to PID variables.}
  \label{overlay}
\end{figure}

\begin{figure}
  \includegraphics[width=\linewidth]{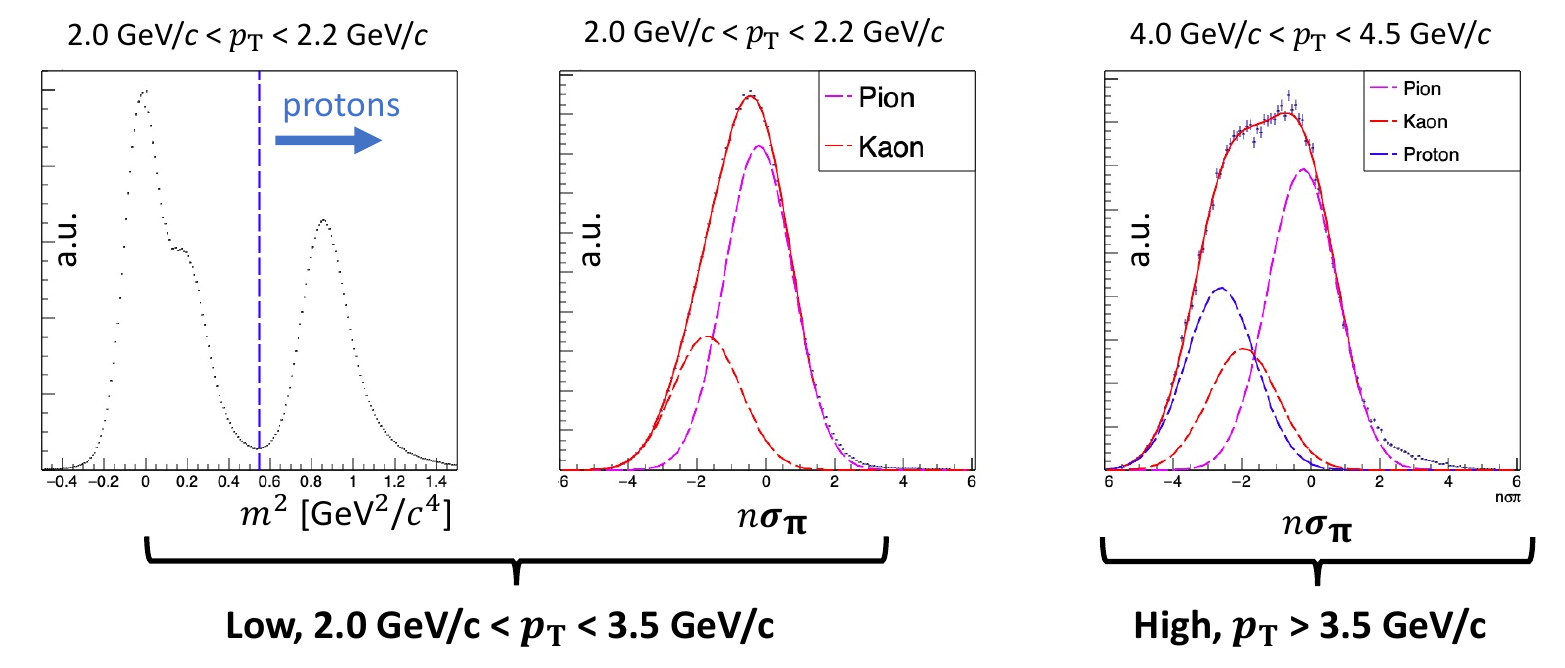}
  \caption{PID technique for low and high $p_{\rm{T}}$ regimes. The left two plots show an example of the method employed in the low $p_{\rm{T}}$ regime; a direct count of protons from $m^{2}$ and a double gaussian fit for pions and kaons in $n\sigma_{\pi}$. The right plot shows the high $p_{\rm{T}}$ regime technique; a triple gaussian fit in $n\sigma_{\pi}$.}
  \label{PID}
\end{figure}

A jet-track correlation technique is employed to find a sample of fully reconstructed jets with tracks identified by the STAR Time-of-Flight (ToF) and Time Projection Chamber (TPC) information to achieve Particle Identification (PID) in jets. Jets are reconstructed using the anti-$k_{\rm{T}}$ algorithm, with various radii parameters $R$, and constituent transverse momentum, $p^{\mathrm{const}}_{\rm{T}}$ , minima~\cite{antikt}. Only the leading jet in each event with a raw jet transverse momentum, $p^{\rm{raw}}_{\rm{T}}$, above $10$ GeV/$c$ is considered. For each event, once an axis for such a jet is determined, correlations are constructed in $\eta$ and $\phi$ for all charged particle tracks in the event, achieving a distribution in $d\eta$ and $d\phi$, where $d\eta = \eta_{\rm{jet}} - \eta_{\rm{track}}$ and $d\phi = \phi_{\rm{jet}} - \phi_{\rm{track}}$. Once this raw correlation signal is constructed, there is a latent geometric structure from pair acceptance that must be corrected. A mixed event method is employed for this correction. The mixed event distribution is constructed by correlating all tracks in an event with a jet axis from a different event of similar centrality and vertex position. The resulting distribution is normalized to unity at maximum. To implement the correction, the signal correlation is divided by the mixed event distribution. The result is a flat underlying event distribution in $d\eta$. Figure~\ref{accept} shows each step of the pair acceptance correction process. \\
\indent After pair acceptance correction is applied, a circular region with a radius equivalent to the anti-$k_{\rm{T}}$ $R$ is selected from correlations as the jet signal, and a region of equal area away from the peak in $d\eta$ is selected as the underlying event (UE). The UE selection is constrained to the same region in $d\phi$ rather than d$\eta$ to ensure an accurate assessment of the hadrochemistry sitting beneath the signal, as it has been shown that baryons and mesons exhibit different flow behavior in azimuth~\cite{flow}. Histograms are constructed in d$\phi$, d$\eta$, $m^{2}$, and $n\sigma_{\pi}$ from these selections, and UE is subtracted from jet in all four parameters. The normalized energy loss is defined as $n\sigma_{\pi} = \frac{\ln{(dE/dx)_{\rm{measured}}}-\ln{(dE/dx)_{\rm{theory}}^{\rm{\pi}}}}{\sigma(\ln{(dE/dx)})}$, where $(dE/dx)_{\rm{measured}}$ is the measured energy loss from the TPC,  $(dE/dx)_{\rm{theory}}^{\rm{\pi}}$ is the theoretical expectation for the energy loss of a pion, and $\sigma(\ln{(dE/dx)})$ is the resolution of the energy loss measurement. Mass squared is measured using $\beta$ from ToF as $m^{2} = p^{2}(1/\beta^2 - 1)$, where $p$ is the measured track momentum. An overlay in d$\phi$ is included in Fig.~\ref{overlay} to show how the UE selection accurately captures the background level beneath the jet peak, as well as an overlay in mass squared as an example of the PID distributions corresponding to the jet and UE regions. \\
\indent Particle identification is achieved using two key parameters: $m^{2}$ derived from ToF information, and energy loss per unit distance, $dE/dx$, derived from TPC tracking information. Two different methods are employed at the low and high $p_{\rm{T}}$ regimes, as shown in Fig.~\ref{PID}. In the low $p_{\rm{T}}$ regime, ToF resolution allows the proton yields to be directly bin-counted, given the clean separation of the proton peak. The remaining bins are then projected onto $n\sigma_{\pi}$ and fit with a double Gaussian to extract the pion yield. In the high $p_{\rm{T}}$ regime, ToF resolution deteriorates. However, at this point, there is improved separation in $dE/dx$ between particle species, so the full distribution is fit with a triple Gaussian to extract both proton and pion yields simultaneously. \\

\section{Results}

\begin{figure}
  \includegraphics[width=\linewidth]{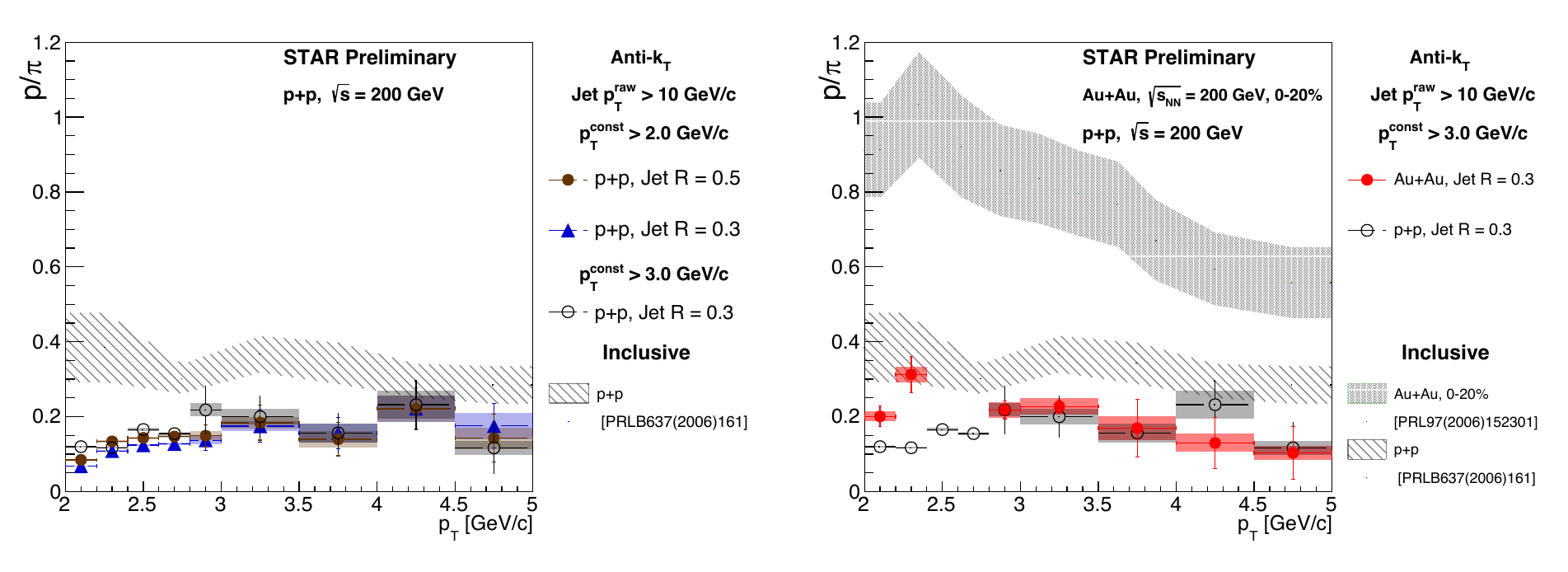}
  \caption{Right: In-jet $p/\pi$ ratios for $p$+$p$ collisions at $\sqrt{s} = 200$ GeV. All jet selections use the anti-$k_{\rm{T}}$ algorithm and have jet $p^{\rm{raw}}_{\rm{T}} > 10$ GeV/$c$. Ratios are shown with R = 0.5 and R = 0.3 for $p^{\rm{const}}_{\rm{T}} > 2.0$ GeV/$c$, and with R = 0.3 for $p^{\rm{const}}_{\rm{T}} > 3.0$ GeV/$c$. The grey band represents the $p/\pi$ ratio from inclusive proton and pion yield measurements~\cite{pp}. Left: In-jet $p/\pi$ ratios for Au+Au collisions at $\sqrt{s_{\mathrm{NN}}} = 200$ GeV, 0-20\% centrality, and $p$+$p$ collisions at $\sqrt{s} = 200$ GeV. All jet selections use the anti-$k_{\rm{T}}$ algorithm and have jet $p^{\rm{raw}}_{\rm{T}} > 10$ GeV/$c$. Ratios are shown with R = 0.3 for $p^{\rm{const}}_{\rm{T}} > 3.0$ GeV/$c$ for both $p$+$p$ and Au+Au samples. Published inclusive $p/\pi$ ratios are included for both $p$+$p$ and Au+Au data~\cite{spectra, pp}.}
  \label{result}
\end{figure}

The $p/\pi$ ratio for jets in $p$+$p$ collisions at $\sqrt{s} = 200$ GeV is shown in Fig.~\ref{result} left for three jet selections: with  $p^{\rm{const}}_{\rm{T}}$ $> 2.0$ GeV/$c$ two anti-$k_{\rm{T}}$ radii are included, $R = 0.3$, and $R = 0.5$. For $p^{\rm{const}}_{\rm{T}} > 3.0$ GeV/$c$ one sample is shown, $R = 0.3$. All jet selections are leading jets with jet $p^{\rm{raw}}_{\rm{T}} > 10$ GeV/$c$. There is a strong preference for pion production over proton production for all jets from $p$+$p$ collisions. The observed in-jet ratio sits below that measured for inclusive $p$+$p$ data, a measure of the full event rather than only within jet, in the low $p_{\rm{T}}$ regime. This feature is not unexpected, as a preliminary report by ALICE has shown similar trends for $p$+$p$ collisions at the LHC~\cite{thesis}. The selection of different radii allows us to study jets with different fragmentation processes. Broader jet selections (for the same $p^{\rm{raw}}_{\rm{T}}$) could have a higher fraction of gluon jets~\cite{JetShape}. This may or may not come into play, given that at RHIC energies, QCD predicts predominantly quark jets~\cite{QCD}. Varying the $p^{\rm{const}}_{\rm{T}}$ minimum provides a control on the hardness of fragmentation within the sample of jets. A higher $p^{\rm{const}}_{\rm{T}}$ minimum corresponds to a harder sample of jets. The results in Fig.~\ref{result} left show that varying either of these variables for jet collections in $p$+$p$ collisions does not impact the resulting in-jet $p/\pi$ ratio.\\
\\
\indent The $p/\pi$ ratio for jets in 0-20\% central Au+Au collisions at $\sqrt{s_{\mathrm{NN}}} = 200$ GeV with $p^{\rm{const}}_{\rm{T}} > 3.0$ GeV/$c$, jet $p^{\rm{raw}}_{\rm{T}} > 10$ GeV/$c$, and $R = 0.3$ are shown in Fig.~\ref{result} right. This ratio shows no significant deviation from the $p/\pi$ ratio in $p$+$p$ for the same jet selection criteria, indicating no significant medium effect on the in-jet $p/\pi$ ratio for this particular jet selection. This selection of leading jets with a narrow anti-$k_{\rm{T}}$ radius and a high $p^{\rm{const}}_{\rm{T}}$ minimum is biased toward the hardest scattering in the event collection. The collection of leading jets in central heavy-ion data may be subject to "survivor bias," that is, have the least interaction with the medium, which could, perhaps, explain the little to no modifications found in $p/\pi$ ratios from jets in Au+Au. In Au+Au data, the $p^{\rm{const}}_{\rm{T}}$ minimum provides a control on how much of the background is picked up by the jet finder algorithm, so lowering this parameter introduces additional background that must be evaluated and corrected. A further extension of this study to larger $R$ and lower $p^{\rm{const}}_{\rm{T}}$ minimum is underway to probe different hardness and fragmentation processes. Given the consistency of the in-jet ratios from $p$+$p$ across various jet selections, any deviation found in Au+Au could be attributed to QGP medium effects.

\section{Conclusion}
The first measurement of in-jet relative baryon-to-meson production from the STAR experiment is presented for $p$+$p$ and central Au+Au collisions at $\sqrt{s_{\mathrm{NN}}} = 200$ GeV. The study of the $p$+$p$ data sample has shown that for multiple jet selections, the $p/\pi$ ratio demonstrates a strong preference for pion over proton production and is indistinguishable within the uncertainties for different jet selections considered. This ratio sits below the published inclusive ratio for $p$+$p$ events of the same energy. Measurements reported for central Au+Au collisions show that anti-$k_{\rm{T}}$ R = $0.3$  jets with constituent $p^{\rm{const}}_{\rm{T}} > 3.0$ GeV/$c$  there exists a similar preference for pion over proton production among jet fragments. The resulting $p/\pi$ ratio hints at no baryon enhancement and is significantly below the ratio reported for inclusive Au+Au collisions of the same centrality. The extension of this analysis to further jet selections in Au+Au is in progress.


\begin{thebibliography}{99}

\bibitem{flow}
A. Adare, et al.,
\emph{Scaling properties of azimuthal anisotropy in Au+Au and Cu+Cu collisions at $\sqrt{s_{\mathrm{NN}}} = 200$ GeV},
\href{https://doi.org/10.48550/arXiv.nucl-ex/0608033}
{\emph{Phys.Rev.Lett.}98:162301,2007}

\bibitem{RAA}
The CMS Collaboration,
\emph{Charged-particle nuclear modification factors in PbPb and pPb collisions at  $\sqrt{s_{\mathrm{NN}}} =  5.02$ TeV},
\href{https://doi.org/10.48550/arXiv.1611.01664}
{\emph{JHEP}04 (2017) 039}

\bibitem{JetQuench}
The CMS Collaboration,
\emph{Decomposing transverse momentum balance contributions for quenched jets in PbPb collisions at $\sqrt{s_{\mathrm{NN}}} = 2.76$ TeV},
\href{https://doi.org/10.48550/arXiv.1609.02466}
{\emph{JHEP}11 (2016) 055}

\bibitem{spectra}
B.I. Abelev, et al.,
\emph{Identified baryon and meson distributions at large transverse momenta from Au+Au collisions at $\sqrt{s_{\mathrm{NN}}} = 200$ GeV },
\href{https://doi.org/10.48550/arXiv.nucl-ex/0606003}
{\emph{Phys.Rev.Lett.}97:152301,2006}

\bibitem{ampt}
A. Lou, et al.,
\emph{Enhancement of baryon-to-meson ratios around jets as a signature of medium response},
\href{https://doi.org/10.1016/j.physletb.2022.137638}
{\emph{Pys.Lett.B}837:137638, 2023}


\bibitem{antikt}
M. Cacciari, G. P. Salam, G. Soyez,
\emph{The anti-$k_{\rm{T}}$ jet clustering algorithm},
\href{https://doi.org/10.1088/1126-6708/2008/04/063}
{\emph{JHEP}04 (2008) 063}

\bibitem{pp}
J. Adams, et al.,
\emph{Identified hadron spectra at large transverse momentum in $p$+$p$ and $d$+Au collisions at $\sqrt{s_{\mathrm{NN}}} = 200$ GeV},
\href{https://doi.org/10.48550/arXiv.nucl-ex/0601033}
{\emph{Phys.Lett.B}637:161-169,2006}


\bibitem{thesis}
B. Andreas Hess,
\emph{Particle Identification in Jets and High-Multiplicity pp Events with the ALICE TPC},
\href{https://doi.org/10.15496/publikation-6879}
{CERN-THESIS-2015-158}

\bibitem{JetShape}
M. Tasevsky,
\emph{Differences between Quark and Gluon jets as seen at LEP},
\href{ https://doi.org/10.48550/arXiv.hep-ex/0110084}
{\emph{OPAL}CR-479}

\bibitem{QCD}
PHENIX,
\emph{An Upgrade Proposal from the PHENIX Collaboration},
\href{https://arxiv.org/pdf/1501.06197.pdf}
{2014}





\end{thebibliography}
\end{document}